\documentclass[prd,preprint,eqsecnum,nofootinbib,preprintnumbers,
tightenlines,longbibliography]{revtex4}
\usepackage{afterpage}
\usepackage{mathrsfs}
\usepackage{graphicx}
\usepackage{bm}
\usepackage{paralist}
\usepackage[colorlinks,linkcolor=blue,citecolor=blue]{hyperref}

\advance\parskip 1.9pt
\advance\voffset -0.2in

\begin{document}

\title{Contributions to the event-by-event charge asymmetry dependence 
for the elliptic flow of $\pi^{+}$ and $\pi^{-}$ in\\heavy-ion collisions}
\author{Adam Bzdak}
\email{abzdak@bnl.gov}
\affiliation{RIKEN BNL Research Center, Brookhaven National Laboratory,\\
Upton, NY 11973, USA}
\author{Piotr Bo\.zek}
\email{piotr.bozek@ifj.edu.pl}
\affiliation{AGH University of Science and Technology, 
Faculty of Physics and Applied Computer Science, PL-30-059 Krak\'ow, Poland
\\ 
Institute of Nuclear Physics, PL-31342 Krak\'ow, Poland
\vspace{15pt} 
}
 
\begin{abstract}
We discuss various contributions to the event-by-event charge-asymmetry
dependence of $\pi^{+}$ and $\pi^{-}$ elliptic flow, recently measured 
by the STAR Collaboration at RHIC.
It is shown that under general assumptions, the difference 
between $v_2^+$ and $v_2^-$ at a given fluctuating value of an
asymmetry parameter, $A$, is a linear function of $A$, as observed in the
preliminary data. 
We discuss two mechanisms that are qualitatively consistent with 
the experimental data and result in a signal of the correct order of magnitude. 
Our subsequent hydrodynamic calculations, assuming local charge 
conservation at freeze-out, yield a qualitative and partial quantitative 
understanding of the observed signal, so offering a detailed test of the 
hydrodynamic model in heavy-ion collisions.
\end{abstract}

\maketitle


\section{Introduction}

Recent experimental measurements of elliptic anisotropy have challenged
the conventional hydrodynamic picture of relativistic heavy-ion collisions
that had proven very effective in facilitating our understanding of many 
features of particle production in such collisions 
\cite{Ollitrault:2012cm,Heinz:2013th,Gale:2013da}.

For instance, as reported in Ref. \cite{Mohanty:2011nm}, at a given energy
and centrality of a collision the elliptic flow for particles and antiparticles
is slightly different. In particular the elliptic flow for protons is larger
than for antiprotons, and the difference between the two decreases with
increasing energy. This effect can be understood either as a result of
baryon stopping \cite{Dunlop:2011cf,Steinheimer:2012bn} or by including the
mean-field potentials in the hadronic phase, as shown in the AMPT model \cite%
{Xu:2012gf}.

An intellectually more challenging experimental observation is the dependence of 
elliptic flow for $\pi ^{+}$ and $\pi ^{-}$ as a function of the event-by-event 
fluctuating asymmetry parameter, 
\begin{equation}
A=\frac{N_{+}-N_{-}}{N_{+}+N_{-}},  \label{A}
\end{equation}%
where $N_{+}$ and $N_{-}$ denote, respectively, the numbers of positively and
negatively charged particles at a given acceptance region. According to the
preliminary STAR data \cite{Ke:2012qb,Wang:2012qs}, at a certain energy and
centrality class, the elliptic flow for $\pi ^{+}$ is a decreasing function of $%
A $, and, for $\pi ^{-}$, the opposite trend is observed. In particular%
\begin{equation}
\left\langle v_{2}^{-}\right\rangle _{A}-\left\langle v_{2}^{+}\right\rangle
_{A}=c+r A,  \label{v2diff}
\end{equation}%
where $\left\langle v_{2}\right\rangle _{A}$ denotes an average elliptic
flow at a given value of $A$, and $c$ and $r$ are parameters. The
preliminary STAR data shows that in Au+Au mid-peripheral collisions at
various energies $r_{\text{exp}}\approx 0.03$ \cite{Ke:2012qb,Wang:2012qs},
where both $A$ and $v_{2}$ are measured in $|\eta |<1$.

As shown in \cite{poster}, this dependence cannot be understood in the baryon-
(or isospin-) stopping scenario. Moreover, both UrQMD \cite{Bleicher:1999xi}
and AMPT \cite{Lin:2004en} models fail to describe the linear relation (\ref%
{v2diff}) \cite{Ke:2012qb}. Recently, an interesting interpretation of 
Eq. (\ref{v2diff}) was proposed as evidence of the effects of a chiral magnetic wave
\cite{Kharzeev:2010gd,Burnier:2011bf,Burnier:2012ae}.  
This effect is closely related to the chiral magnetic effect \cite{Kharzeev:2007jp},
discussed extensively in the recent literature, e.g., the recent review 
in \cite{Bzdak:2012ia}.

In this paper, we demonstrate that Eq. (\ref{v2diff}) is consistent with the
hydrodynamic picture of heavy-ion collisions with local charge conservation
at freeze-out. We discuss the relation between Eq. (\ref{v2diff}) and
the physics underlying the charge balance function \cite{Jeon:2001ue}, successfully
described by hydrodynamic models, e.g., \cite%
{Schlichting:2010qia,Bozek:2012en}. In the following section, we derive a general relation
between $\langle v_{2}^{\pm }\rangle _{A}$ and $A$, yielding a linear dependence. 
In section $3$, we discuss two mechanisms
that qualitatively and partly quantitatively reproduce the experimental
data. In section $4$, we detail state-of-the-art $3+1$-dimensional
hydrodynamic calculations with local charge conservation, observing
a strong signal consistent with the experimental one, within a factor of $2$.
Section $5$ gives our comments and conclusions.

\section{General relations}

In this section we derive the relation between elliptic flow at a given 
$A$, $\left\langle v_{2}\right\rangle _{A}$, as a
function of $\left\langle v_{2}\right\rangle _{A=0}$ and $A$. Here, for
simplicity, we assume that the distribution of $A$ at a given centrality
class is symmetric with respect to $A=0$.\footnote{%
So that we can expand all variables around $A=0$.}

Straightforwardly, we notice that for a sufficiently small $A$%
\begin{equation}
\left\langle N_{+}-N_{-}\right\rangle _{A}=\alpha A\left\langle
N_{+}+N_{-}\right\rangle _{A=0}=2\alpha A\left\langle N_{+}\right\rangle
_{A=0},
\end{equation}%
where $\alpha $ is a parameter whose value is close to unity\footnote{%
If the same particles are used to calculate $A$ and $\left\langle N_{\pm
}\right\rangle _{A}$, then $\alpha \simeq 1$. In the STAR measurement, $%
\left\langle N_{\pm }\right\rangle _{A}$ and $A$ are calculated from similar
sets of particles, but not identical ones. This distinction may engender a value 
that differs slightly from $1$.}, see Eq. (\ref{A}). Here, $\left\langle .\right\rangle
_{A=0} $ denotes an average over all events characterized by $A=0$. 
We readily notice that
\begin{equation}
\left\langle N_{+}+N_{-}\right\rangle _{A}=2\left\langle N_{+}\right\rangle
_{A=0}\left( 1-\beta ^{2}A^{2}+...\right) .
\end{equation}%
The left-hand side of the above equation is invariant under the transformation $%
A\rightarrow -A$; thus, only the even powers of $A$ are allowed.\footnote{%
To obtain $A=0.1$, for example, we could increase $N_{+}-N_{-}$, or/and decrease $%
N_{+}+N_{-}$. Thus, for larger values of $A$, $N_{+}+N_{-}$ is expected to
decline as a function of $A$. We verified this in our simple model
discussed in the next section, where $\beta \approx 2.3$, and $\alpha =0.96$.}
These equations above support our evaluation of $\left\langle N_{+}\right\rangle _{A}$
and $\left\langle N_{-}\right\rangle _{A}$%
\begin{equation}
\left\langle N_{\pm }\right\rangle _{A}=\left\langle N_{+}\right\rangle
_{A=0}\left( 1\pm \alpha A-\beta ^{2}A^{2}+...\right) .  \label{Npm}
\end{equation}

To express elliptic flow at a given $A$ as a function of $A$ we write%
\begin{equation}
\left\langle \frac{d^{3}N_{+}}{d\varphi d^{2}p_{t}}\right\rangle
_{A}=\left\langle \frac{d^{3}N_{+}}{d\varphi d^{2}p_{t}}\right\rangle
_{A=0}+\left\langle \frac{d^{3}N_{+}}{d\varphi d^{2}p_{t}}\right\rangle _{%
\text{asym}},  \label{d3N}
\end{equation}%
and analogously for negative particles. Here, $\left\langle .\right\rangle _{%
\text{asym}}$ represents an average number of particles responsible for a
non-zero value of $A$. Integrating the above equation over $\varphi $ and $p_{t}$
we obtain Eq. (\ref{Npm}) that allows us to express $\left\langle
N_{+}\right\rangle _{\text{asym}}$ by $\left\langle N_{+}\right\rangle
_{A=0} $ and $A$. Multiplying both sides of Eq.~(\ref{d3N}) by $\cos
(2\varphi )$ and integrating over an available phase space, assuming $A<<1$
such that $(1\pm A)^{-1}\approx 1\mp A$ and neglecting the $A^{2}$ terms, we obtain%
\begin{equation}
\left\langle v_{2}^{+}\right\rangle _{A}\approx \left\langle
v_{2}^{+}\right\rangle _{A=0}-\alpha A\left[ \left\langle
v_{2}^{+}\right\rangle _{A=0}-\left\langle v_{2}^{+}\right\rangle _{\text{%
asym}}\right] ,
\end{equation}%
and 
\begin{equation}
\left\langle v_{2}^{-}\right\rangle _{A}\approx \left\langle
v_{2}^{-}\right\rangle _{A=0}+\alpha A\left[ \left\langle
v_{2}^{-}\right\rangle _{A=0}-\left\langle v_{2}^{-}\right\rangle _{\text{%
asym}}\right] ,
\end{equation}%
where $\langle v_{2}^{\pm }\rangle _{A}$ denotes elliptic flow at a given
value of $A$, and $\langle v_{2}^{\pm }\rangle _{\text{asym}}$ represents an
average elliptic flow of particles responsible for a non-zero value of $A$.
The elliptic flow for all events with $A=0$ is denoted by $\langle v_{2}^{\pm
}\rangle _{A=0}$.

As is evident from the above equations, a linear dependence on $A$ is expected 
naturally only if the elliptic flow of particles contributing to a non-zero
asymmetry differs from the elliptic flow of particles leading to $A=0$. In
the next section, we discuss two mechanisms that naturally lead to 
$\langle v_{2}\rangle _{A=0}>\langle v_{2}\rangle _{\text{asym}}$, being
qualitatively consistent with the preliminary data.

\section{Two mechanisms}

In this section, we discuss two mechanisms that offer qualitative and
partly quantitative agreement with the preliminary STAR data.

From various measurements of the balance function in rapidity, we know that
electric charges balance each other with a characteristic distance
between the charges of about $0.5$ unit of rapidity \cite{Aggarwal:2010ya}. 
This finding is confirmed successfully in the hydrodynamic model with locally
conserved charge production at freeze-out \cite%
{Schlichting:2010qia,Bozek:2012en}.\footnote{%
In contrast to models with an initial charge creation (such as UrQMD) that cannot
describe the experimental data on the balance function \cite{Aggarwal:2010ya}.} 
Consequently, a pair of pions with opposite charges
originating from a fluid element or a resonance, create a non-zero $A$ in a
given rapidity window provided that one pion is produced inside of this window and
the other outside it. If both charges are produced inside a bin they do not 
contribute to $N_{+}-N_{-}$. In the STAR experiment, the elliptic flow is
measured in $\left| \eta \right| <1$ that is broader than the
typical width of the balance function.

\begin{figure}[h]
\includegraphics[scale=0.9]{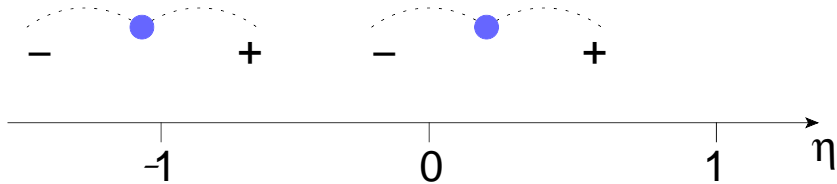}
\caption{Clusters (resonances, fluid elements), denoted by the blue dots,
positioned close to $\eta=0$, decay into final particles located inside the
rapidity bin $\left| \eta\right| <1$ and clearly cannot create a non-zero
asymmetry, $A$. Only clusters located close to the boundary can produce one
pion inside and one pion outside of our rapidity bin. From the PHOBOS data,
we know that elliptic flow is slightly smaller away from $\eta=0$.
Consequently, the elliptic flow of particles creating the asymmetry, $A\neq 0$, is
smaller than that of particles leading to zero asymmetry, $A=0$.}
\label{fig_eta}
\end{figure}

In Fig. \ref{fig_eta}, we schematically depict the first mechanism, leading 
to Eq. (\ref{v2diff}). Fluid elements and resonances, in short
clusters, located close to $\eta=0$ usually produce both pions inside the
pseudorapidity window. Only clusters located close to $|\eta|=1$ typically create a
non-zero asymmetry parameter. From the PHOBOS data \cite{Alver:2006wh}, we
know that elliptic flow of charged particles as a function of pseudorapidity 
can be roughly described as $v_{2}(\eta )\sim 1-|\eta |/6$  
(further discussed in section 5).
Thus, clusters located close to $|\eta|=1$ have smaller
elliptic flow compared to clusters located close to $\eta=0$.
Consequently, the elliptic flow of pairs responsible for a non-zero asymmetry is
smaller than that of pairs leading to $A=0$. Using Eq. (%
\ref{v2diff}) we conclude that $\langle v_{2}^{+}\rangle _{A}$ decreases and 
$\langle v_{2}^{-}\rangle _{A}$ increases as a function of $A$.\footnote{By selecting 
events with $A>0$, for example, we trigger configurations
wherein positive pions are inside a rapidity window and the negative ones are outside it.
In this way, we add positive particles with small $v_{2}$ and, effectively, 
the $v_{2}$ for positive particles is lowered. Since the number of negative particles
with a small $v_{2}$ is reduced, thus effectively, their $v_{2}$ is increased.}

\vspace{7pt}

\begin{figure}[h]
\includegraphics[scale=0.9]{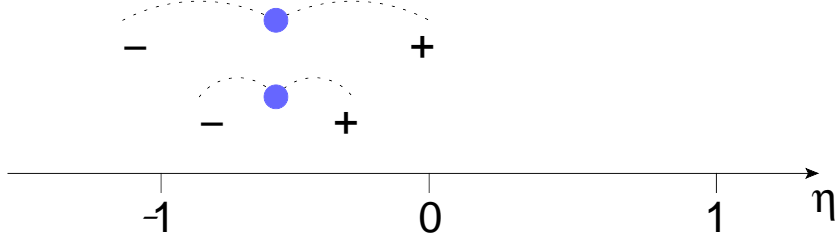}
\caption{Clusters (resonances, fluid elements), denoted by the blue dots,
with low transverse momentum (velocity) produce particles separated more in
rapidity compared to clusters with high transverse momentum.
Consequently, non-zero values of $A$ are mainly driven by clusters with a small 
$p_{t}$ that are characterized by a small $v_{2}$. Thus, the elliptic flow of
particles creating asymmetry, $A\neq 0$, is smaller than that of
particles leading to zero asymmetry, $A=0$.}
\label{fig_pt}
\end{figure}

The second mechanism is explicitly related to the late locally conserved
charge production in the hydrodynamic picture, see Fig. \ref{fig_pt}.
Clusters with small transverse momentum (resonances) or small transverse
velocity (fluid elements) usually decay into pairs of pions with a large
rapidity separation (a broad balance function). In contrast, clusters with
large transverse momentum or velocity engender a small rapidity separation
(a narrow balance function). Undoubtedly, clusters with a broad balance function
are more likely to produce a non-zero charge asymmetry parameter, $A$. On the
other hand, we know that the elliptic flow of clusters with a small transverse 
momentum (velocity) is less than that of clusters with a high
transverse momentum, e.g., \cite{Gale:2013da}. Consequently $%
\left\langle v_{2}\right\rangle _{\text{asym}}$ is smaller than $%
\left\langle v_{2}\right\rangle _{A=0}$, so leading to the linear dependence of $%
\langle v_{2}^{-}\rangle _{A}-\langle v_{2}^{+}\rangle_{A}$ on $A$.

\vspace{7pt}

To quantify these effects, we undertook
a simplified Monte Carlo calculation.\footnote{%
In each event, we sampled clusters from the Poisson
distribution with an average number of clusters, $\left\langle
N_{c}\right\rangle =570$. All clusters were uniformly distributed in
pseudorapidity over $|\eta |<3$ and in transverse momentum according to the
thermal distribution, $e^{-2p_{t}/\left\langle p_{t}\right\rangle }$ with $%
\left\langle p_{t}\right\rangle =0.7$ GeV. A cluster with a given $\eta $
and $p_{t}$ decays into two oppositely charged particles, located
symmetrically around $\eta $. The pseudorapidity distance between them is
sampled from the Gaussian distribution with its standard deviation $\sigma
=1/(1+p_{t}/\left\langle p_{t}\right\rangle )$, so that for a cluster with $%
p_{t}=\left\langle p_{t}\right\rangle $, $\sigma =0.5$. This dependence on $%
p_{t}$ approximately agrees with a rho meson decay at a given $p_{t}$ into
two pions. Finally, we sampled the azimuthal angles of the two particles using $%
1+2v_{2}(p_{t},\eta )\cos (2\varphi )$ with $v_{2}(p_{t},\eta )$ given by $%
0.04(1-\left| \eta \right| /6)p_{t}/\left\langle p_{t}\right\rangle $ for $%
p_{t}<2$ GeV, and $0.04(1-\left| \eta \right| /6)2/\left\langle
p_{t}\right\rangle $ for $p_{t}>2$ GeV. $A$ and $v_{2}$ are calculated for
particles in $|\eta |<1$. In our calculation we obtained $\left\langle
v_{2}\right\rangle \approx 0.036$, and approximately $380$ charged particles
in the midrapidity region $|\eta |<1$, approximately agreeing with $200$
GeV Au+Au collisions in the $30-40\%$ centrality class.} As shown in Fig. \ref%
{fig_model}, the signal we obtained is of the same order-of-magnitude as the
data. In fact, we obtained $r\approx 0.02$, see Eq. (\ref{v2diff}), that is
only a factor of $1.5$ smaller than the preliminary STAR data, $r_{\text{exp}%
}\approx 0.03$.  
In our calculation, both $A$ and $v_{2}$ are calculated for particles in $|\eta |<1$. 
However, we note that our results represent only
an order-of-magnitude estimate, and more advanced calculations are required
to draw definite conclusions. 
\begin{figure}[h]
\includegraphics[scale=0.48]{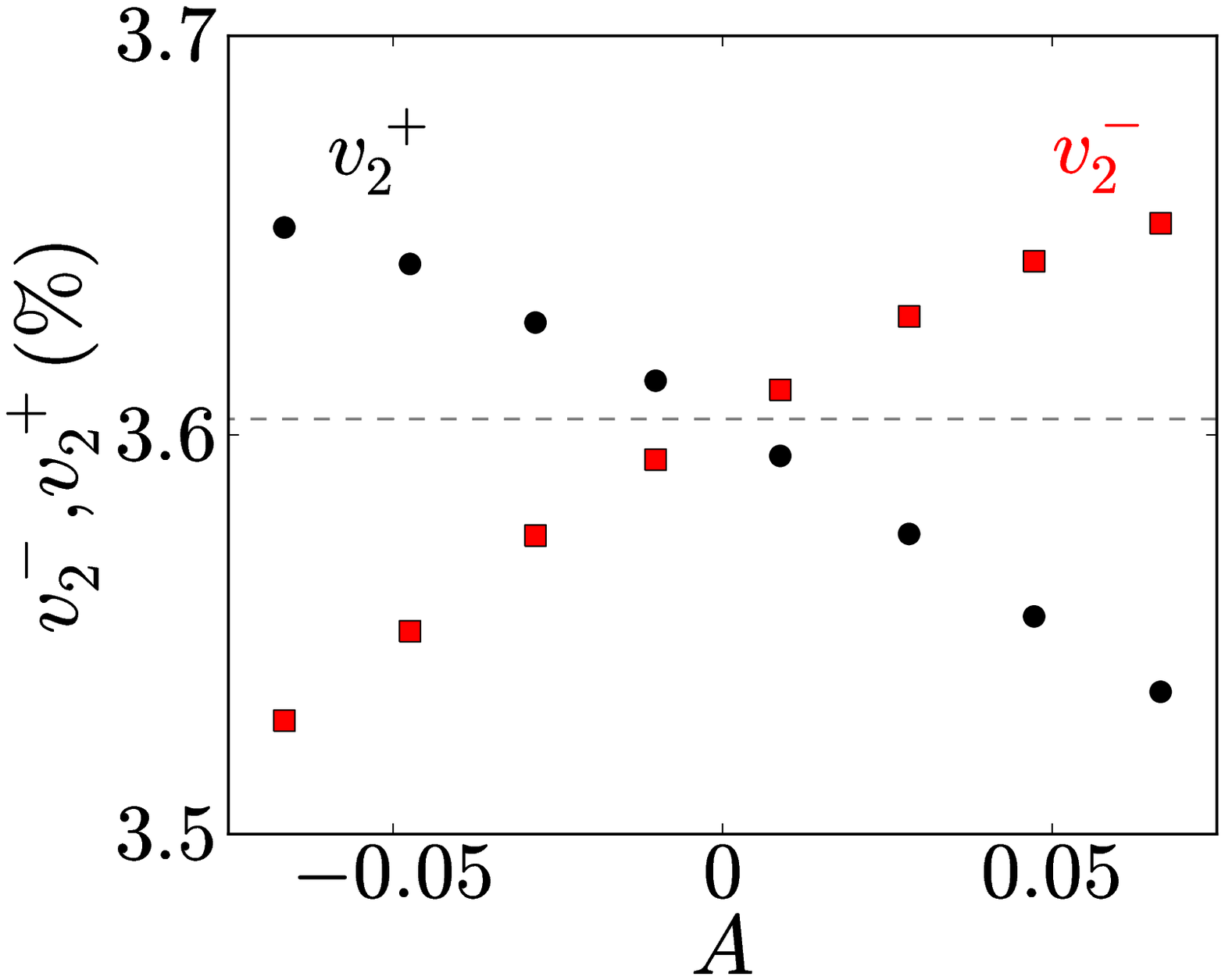} \hspace{0.3cm} %
\includegraphics[scale=0.48]{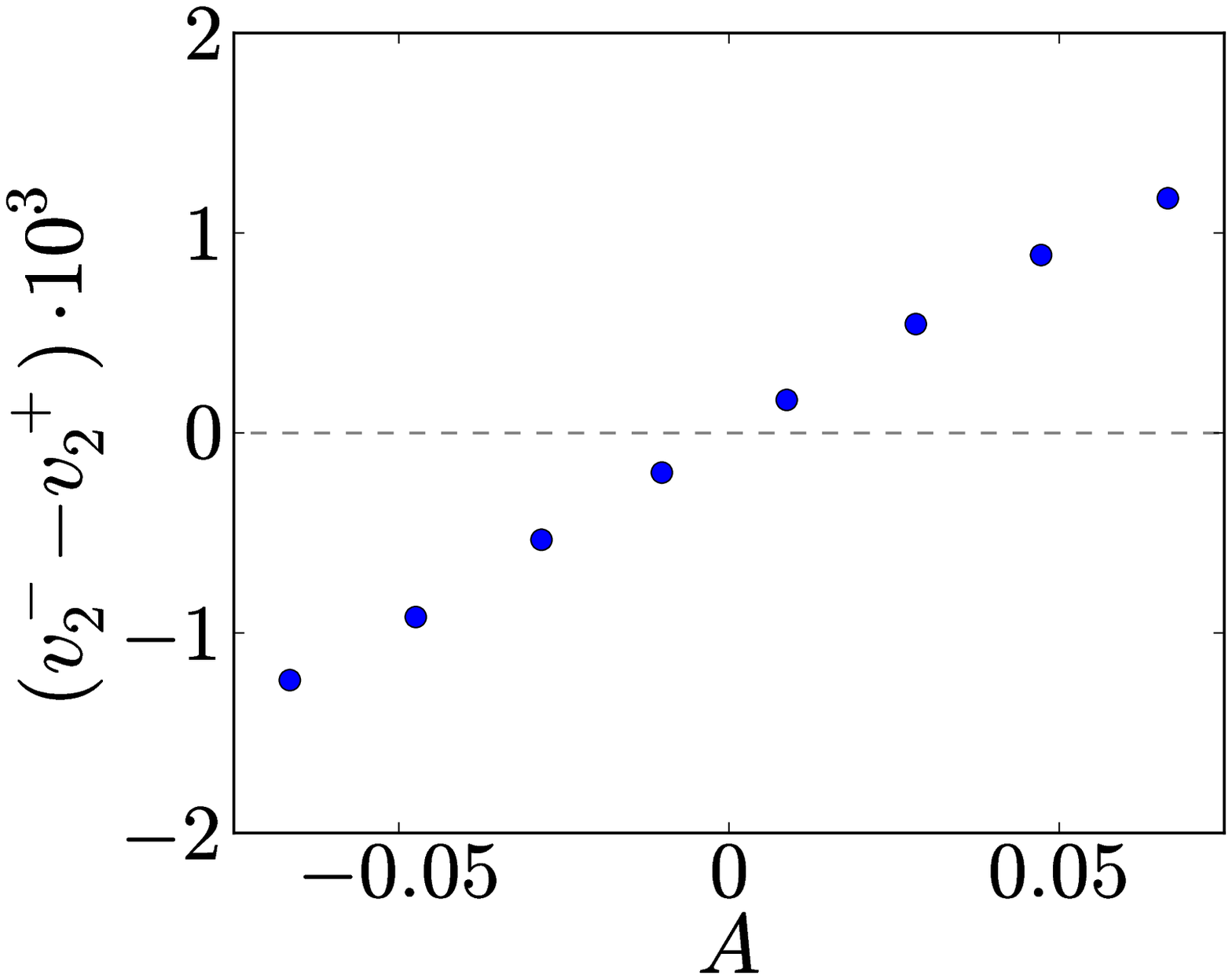}
\caption{Our results for the elliptic flow coefficient for positive and 
negative particles
as a function of asymmetry parameter $A=(N_{+}-N_{-})/(N_{+}+N_{-})$. In our
simplified model we obtain $r\approx 0.02,$ see Eq. (\ref{v2diff}), to be
compared with the preliminary STAR data $r_{\text{exp}}\approx 0.03$. Note
that our results represent only an order-of-magnitude estimate.}
\label{fig_model}
\end{figure}

We checked that the effect schematically presented in Fig. \ref{fig_eta} is responsible
approximately for $35\%$ of the signal shown in Fig. \ref%
{fig_model}. 
We also checked that calculating $v_{2}$ in $|\eta |<0.2$ and $A$ in $|\eta
|<1$ reduces our final results by about $10\%$. However, when calculating $%
v_{2}$ in $|\eta |<1.5$, and $A$ in $|\eta |<1$, we obtained a significantly 
smaller value, $r\approx 0.006$.
Taking $v_{2}$ in $|\eta |<2$ and $A$ in $|\eta |<1$ lowers our results 
approximately by a factor of $10$.
Finally, we mention that in our simplified analysis we neglected experimental 
cuts in the transverse momentum which may slightly modify our numbers.

It would be desirable to perform a more realistic calculations in an
event-by-event $3+1$-dimensional hydrodynamic model, wherein the effect
presented in Fig. \ref{fig_pt} is naturally present. This problem is
discussed in the next section.

\section{Hydrodynamic calculation}

The collective flow in heavy-ion collisions can be reproduced satisfactory
within relativistic hydrodynamics \cite{Ollitrault:2012cm,Heinz:2013th,Gale:2013da}. 
Statistical emission and
resonance decay at freeze-out account for part of the observed charge
balancing correlations. Additionally, local charge balancing in particle
production \cite{Jeon:2001ue} can be included in the event generator \cite%
{Bozek:2012en}, yielding the right qualitative description of the one- and
two-dimensional charge dependent correlations. We used a $3+1$-dimensional
hydrodynamic model with statistical emission at freeze-out including the local
charge conservation mechanism and resonance decays \cite{Bozek:2011ua}. In
order to study non-flow correlations, particle emission from the freeze-out
hypersurface should be done via Monte-Carlo event generation. The default
implementation of the statistical emission in THERMINATOR \cite{Chojnacki:2011hb}
generates primordial particles with the Poisson distribution for each type of particle. 
In the present calculation, we fix the total charge produced in the
fireball, for simplicity fixing the multiplicities for all particles emitted
from the fireball in each event. The charge asymmetry measured in a limited
acceptance window mainly is due to the separation of an unlike charged pair
as described in the previous section.

\begin{figure}[h]
\includegraphics[scale=0.5]{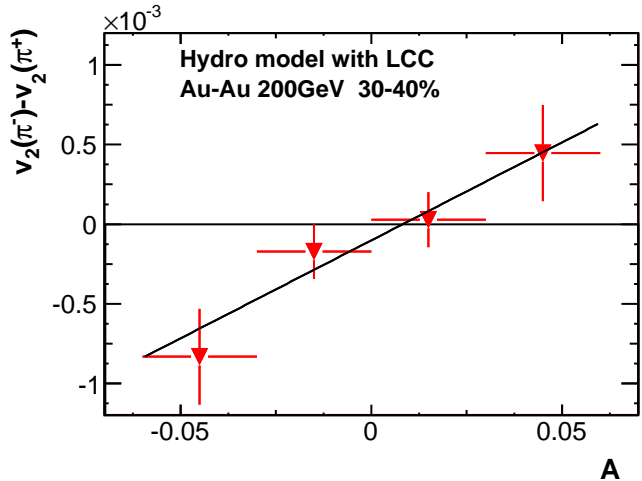}
\caption{The charge asymmetry dependence of $\protect\pi ^{+}$ and $\protect%
\pi ^{-}$ elliptic flow coefficients in the hydrodynamic model followed by
statistical emission with local charge conservation. 
We obtained $r=0.012\pm 0.004$ compared with the preliminary STAR data,
$r_{\text{exp}}\approx 0.03$. }
\label{fig:hydcmw}
\end{figure}

The statistical emission code \cite{Chojnacki:2011hb} follows resonance
decays, and such charge correlations are realistically described in the
model. Together with the effects of local charge conservation, correlating the
emission of primordial particles, it yields a good quantitative description
of charge balance functions in semi-central Au-Au collisions \cite%
{Bozek:2011ua}. To study the weak charge splitting signal, a large statistics is required. 
We generated $10^6$ events for the centrality 30-40\%
wherein the measured signal is the strongest, and where the hydrodynamic model agrees best with 
the data on charge balance functions.
The elliptic flow, $v_{2}$, for $%
\pi ^{+}$ and $\pi ^{-}$ is calculated for events with different charge asymmetry
in the interval $|\eta |<1$, using two subevents with a pseudorapidity
separation $|\Delta \eta | >0.3$ to define $v_2$, following the experimental procedure \cite%
{Ke:2012qb}. The Monte-Carlo calculation with local charge conservation and
resonance decays exhibits the expected signal, viz., a decrease of $v_{2}(\pi
^{+})$ and an increase of $v_{2}(\pi ^{-})$ with the charge asymmetry $A$
in the event (Fig. \ref{fig:hydcmw}). The slope parameter, $r\simeq 0.012$, is
about one half of the observed magnitude. We note that the
pseudorapidity dependence of the elliptic flow $v_{2}(\eta )$ observed
experimentally \cite{Alver:2006wh} is noticeably underestimated in $3+1$%
-dimensional hydrodynamic simulations \cite{Schenke:2011bn,Bozek:2011ua}.
The steeper pseudorapidity dependence of $v_{2}$ is expected to improve the
agreement between the calculated charge splitting of the elliptic flow with
the experimental findings. In the next section, we comment further on this issue. 

Finally, we discuss the centrality dependence of the signal.
Two effects that are not well controlled in hydrodynamic models may cause the decrease 
of the slope parameter for peripheral events. First, the balance function in pseudorapidity is
widening for peripheral events. Second, for peripheral events,
correlations due to elliptic flow are diluted by the larger contribution of
particles emitted from the nonthermal corona in the interaction region,
while the charge asymmetry acquires contributions both from the core and the corona.

\section{Comments and conclusions}

Several comments are warranted.

(i) In this paper we discussed two mechanisms leading to elliptic flow
splitting as a function of the asymmetry parameter, $A$. We note
that we expect a similar effect for higher harmonics, $v_{n}$. In particular,
for the triangular flow%
\begin{equation}
\left\langle v_{3}^{-}\right\rangle _{A} - \left\langle v_{3}^{+}\right\rangle _{A}\sim A,
\end{equation}%
although the signal is expected to be significantly weaker in comparison to the case for $%
v_{2}$. We expect that $r$, see Eq. (\ref{v2diff}), should be smaller
roughly by a factor of $\left\langle v_{2}\right\rangle /\left\langle
v_{3}\right\rangle \approx 3$ \cite{Adare:2011tg}. 
This problem will be discussed elsewhere.

(ii) The key ingredient in our analysis is the assumption of local charge
conservation at freeze-out. In this scenario, particle pairs with higher
momentum are more strongly collimated in rapidity than are pairs with smaller
momentum, see Fig. \ref{fig_pt}. This allows us to understand the
centrality dependence of the rapidity balance function \cite{Aggarwal:2010ya}, 
in contrast to models with initial charge creation (such as UrQMD).
Consequently, in models without late local charge conservation we
do not expect to observe the elliptic flow splitting discussed in this paper.

(iii) According to the PHOBOS data, the elliptic flow of charged particles in the
midrapidity region changes significantly as a function of pseudorapidity %
\cite{Alver:2006wh}. However, the results of the STAR Collaboration indicate
that elliptic flow in the midrapidity region weakly depends on $\eta $ \cite%
{Abelev:2008ae}, so being consistent with hydrodynamic calculations (see
section $4$). Consequently, the mechanism presented in Fig. \ref{fig_eta}
may be suppressed if we take the STAR data into account. However, this is not a
problem because, as we discussed in section $3$, this mechanism is
responsible for only $25\%$ of the measured signal, and $35\%$ of the signal
presented in Fig.~\ref{fig_model}.

\bigskip

In conclusion, we studied the dependence of the elliptic flow coefficients for
positive and negative particles as a function of the event-by-event charge asymmetry
parameter. Recently, this phenomenon was interpreted as evidence of the
chiral magnetic wave. In this paper, we argued that the origin of this effect
may be less exotic and actually is consistent with the hydrodynamic picture
of heavy-ion collisions with late local charge conservation.

We argued that particle pairs leading to a non-zero asymmetry parameter, $%
A\neq 0$, are characterized by a smaller elliptic flow in contrast
to particle pairs resulting in $A=0$. We showed that this fact alone is
sufficient to qualitatively explain the preliminary STAR data, in
particular, a linear dependence of $\langle v_{2}^{-}\rangle
_{A}-\langle v_{2}^{+}\rangle _{A}$, as a function of $A$. Our
quantitative results, based on a simplified Monte Carlo estimation and the
state of the art $3+1$-dimensional hydrodynamic model calculations, are
in agreement, within a factor of $2$, with the preliminary STAR data.

\vspace{10pt} 
\noindent \textbf{Acknowledgments}
\newline
{}
\newline 
We thank H.-U. Yee for interesting discussions. Remarks by V. Koch and 
L. McLerran are highly appreciated. W. Broniowski graciously helped us in modifying 
the THERMINATOR code. A.B. is supported through the RIKEN-BNL
Research Center. The work is partly supported by the National Science Centre, 
Poland, grant DEC-2012/05/B/ST2/02528.

\end{document}